# Superluminal signal conversion in stimulated Brillouin scattering via an optical fiber ring resonator

Liang Zhang, Li Zhan*, Jinmei Liu, Gaomeng Wang, Fangying Tao, Taohu Xu,Qishun Shen

*Department of Physics, Key Laboratory for Laser Plasmas (Ministry of Education), State Key Lab of Advanced Optical Communication Systems and Networks, Shanghai Jiao Tong University, Shanghai, 200240, China*

**Corresponding author, Email: lizhan@sjtu.edu.cn*

We report the superluminal phenomenon of both Stokes and pump light in stimulated Brillouin scattering (SBS) via an optical-fiber ring lasing resonator. In our experiment, the superluminal generation of Stokes light firstly delayed with 33.79 ns and then propagated with the advancement of 93.34 ns within an 8-m single mode fiber (SMF), when the pump power increased from the SBS threshold to a higher power. This proves the first evidence that the optical interaction is determined by the group velocity even at the negative group velocity superluminal propagation. It is important that, this implies the possibility of superluminal information interchange because the results also indicate that the signal conversion between different wavelengths can be realized at the negative group velocity propagation.

To date, many experiments have demonstrated the possibility to control the group velocity of light, resulting in slow, fast or even superluminal light [1-2]. Slow light has been demonstrated with very small group velocity [3-5] and even stopped light [6-7]. Meanwhile, superluminal or even negative group velocity propagation has been demonstrated by recent experiments [8-10]. Here, the anomalous dispersion is generated to result in light propagation with the negative group velocity in material. In such cases, the light pulse appears to exit the medium before entering, which seems at odds with causality. However, a direct consequence of classical interference between light pulse's different frequency components results in the observed superluminal propagation in an anomalous dispersion region [11-12].

Slow/fast light is initially demonstrated by exploiting narrow spectral gain/loss resonances, typically created by electromagnetically induced transparency [4] or coherent population oscillation [9]. Within the gain/loss resonance, the light pulse experiences the normal/anomalous dispersion, which complies with the Kramers-Kronig relations [13]. The group velocity $\upsilon_g = c/n_g = c/(n+\omega dn/d\omega)$ ($c$ is the light speed in vacuum; $n$ is the refractive index; a positive or negative value of $dn/d\omega$ is referred to as normal or anomalous dispersion.) can be smaller/larger than the phase velocity $c/n$ in medium. Thus, superluminal light appears when the group velocity exceeds $c$ if the anomalous dispersion is large enough. Generally, gain resonance results in normal dispersion and slow-light effect, and the depleted wave undergoes loss resonance and fast-light effect.

The SBS approach has been proved as a flexible mechanism to realize the fast/slow light in optical fibers [3, 14-16]. Within the gain band at the Stokes frequency, the light signal is slowed and its delay agumentes as the pump power increases. On the contrary, the attenuated pump light experiences fast-light effect. The possibility of superluminal propagation via SBS has been demonstrated in fibers [17-19], but the light usually experiences tremendous absorption. However, our recent work demonstrated the low-loss superluminal propagation in fibers by employing the Brillouin lasing cavity, which provides a simple but practical platform for studying the superluminal physics [20]. In all previous experiments, the Stokes wave in SBS process exhibits slow light and the pump wave shows fast light. This is the certain conclusion from the Kramers-Kronig relations.

Usually, superluminal effect leads to the controversies of the definition of the information velocity and relativistic causality [21-22]. However, recent experiments demonstrated that the superluminal transfer of information is consistent with the concept of Einstein causality [23-24].It has been demonstrated that the peak of the pulse does propagate backward inside the fiber when the pulse propagates with superluminal or even negative group velocity, whereas the energy flow is in the forward direction [17]. Although many special properties surrounding the superluminal effect have been studied, the signal conversion has not yet been experimentally demonstrated under the superluminal propagation. Especially, a fundamental question still should be addressed: Can the light signal convert to the other wavelength at the negative group velocity propagation? In this case, the energy flow passes in the opposite direction to the signal.

In the present work, we show that the Stokes light is generated with superluminal propagation via a fiber ring Brillouin resonator. Thanks to the powerful enhancement of lased Stokes wave, the pump appears low-loss superluminal propagation. Such large advancement of the group velocity, in turn, makes an ignorable influence towards the Stokes light. The results firstly demonstrated the optical interaction between the pump and Stokes wave is not definitely determined by the phase velocity but by the group velocity, which is still consistent with superluminal phenomenon even at negative group velocity.

SBS is a well known nonlinear effect. When pump light with the frequency $\omega_p$ propagates in a fiber, Stokes light is backward generated with a downshifted frequency $\omega_s = \omega_p - \Omega_B$ ($\Omega_B$ is the Stokes shift). The pump experiences an absorption and anomalous dispersion while the Stokes wave experiences a gain and normal dispersion, which results in fast/slow light effect. With the slowly varying envelop approximation, the group index via SBS is given by[25]

$$n_g = n_{g0} + \left(\frac{g_p c P_p}{\Gamma_B A_{eff}}\right)\frac{1-\delta^2}{(1+\delta^2)^2}, \quad (1)$$

where, $n_{g0}$ is the group index in the absence of SBS gain, $g_p$ is the peak value of gain, $\Gamma_B$ is the gain linewidth, $P_p$ is the pump power, $A_{eff}$ is the effective core area of fiber, and $\delta = 2(\omega_p - \omega_s - \Omega_B) / \Gamma_B$ is the normalized detuning.

After propagating the distance *L*, slow/fast light can be determined by considering the relative advancement $\Delta T = Ln_g / c - Ln / c$. $\Delta T$ =0 if no slow/fast light effect. It performs slow light if $\Delta T < 0$. Otherwise, fast light appears if $\Delta T$ is positive. Negative group velocity propagation appears when $\Delta T$ is larger than $Ln / c$, which seems that the light has exited in the medium before its entering.

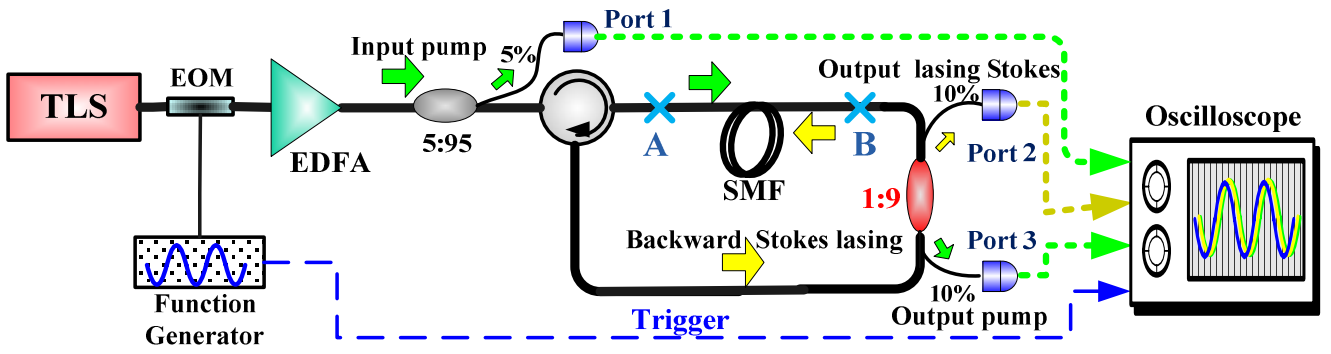

FIG. 1. Experimental setup to observe the superluminal effect of Stokes light via a fiber ring Brillouin resonator. A tunable laser source (TLS) modulated by an electro-optic modulator (EOM) is to produce a sinusoidal pump at 1550 nm. The sinusoidal signal is set at 1-MHz frequency by adjusting the function generator. After boosted by an erbium-doped fiber amplifier (EDFA), the pump light is injected into the SMF to generate SBS and create the backward Stokes wave. Here, 90% of Stokes light is circulated for lasing.

In experimental setup (Fig. 1), we choose a 10-m SMF as the SBS medium in the ring lasing resonator. 90% of backward Stokes light is served to produce SBS-induced loss resonance of the pump light, and 10% power as the output lasing power is detected at port 2. The output pump light is observed at port 3. The temporal traces of the pump light are monitored by an oscilloscope with photo detectors.

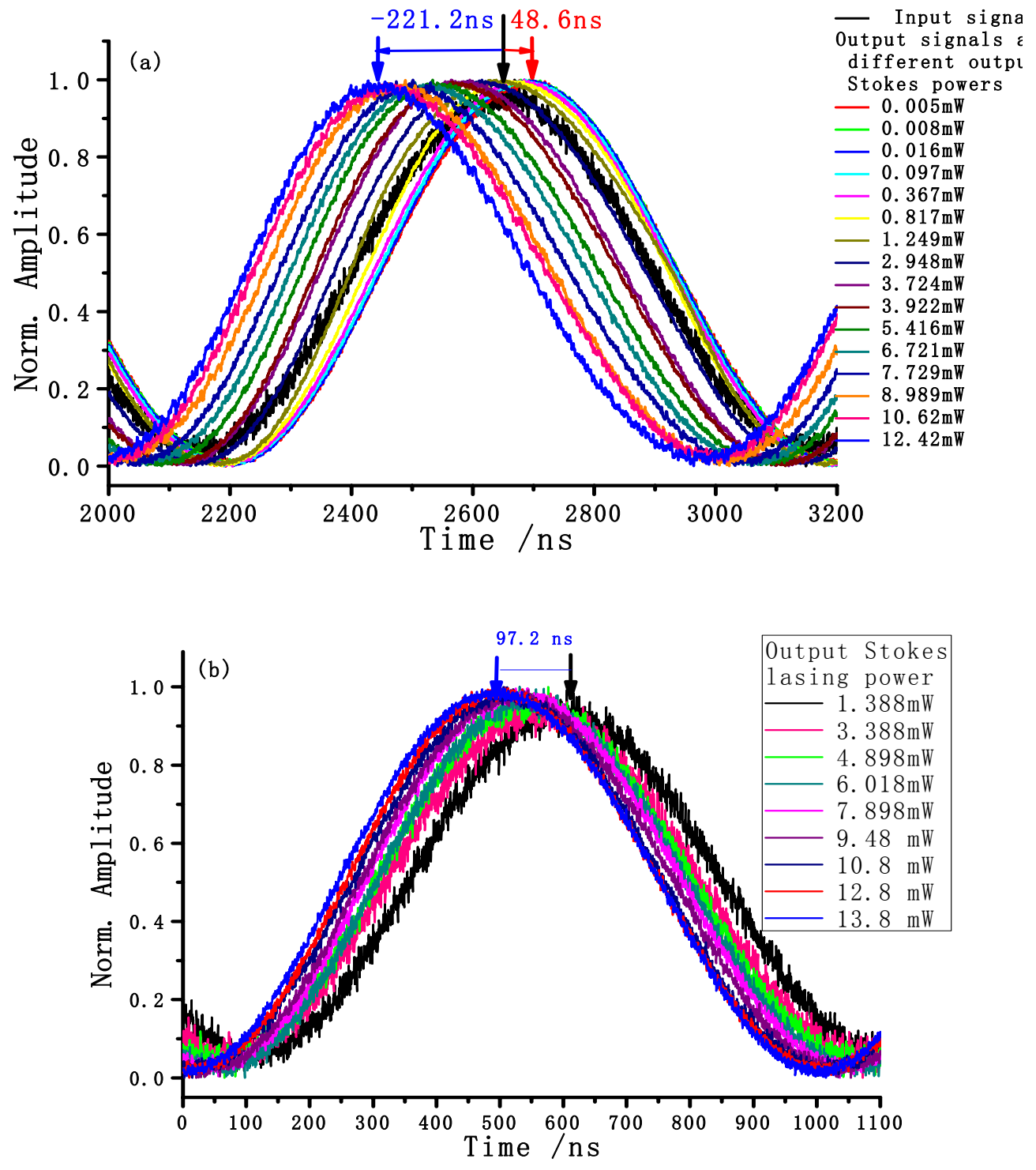

FIG. 2. (a)Waveforms of input signal and output pump signal after propagating through the 10-m SMF, (b)Waveforms of generated Stokes wave after anticlockwise passing through the SMF, for different output Stokes lasing powers.

As shown in Fig. 2, the pump light is modulated to a sinusoidal signal of 1 MHz by the EOM. When the input power is below the SBS threshold, the delay of pump wave is 48.6 ns after passing through 10-m SMF. Superluminal propagation of pump wave appears when the input power reached 338.2 mW. However, in Fig. 2(b), it is surprised that the Stokes light propagates with advancement as the Stokes lasing power increased. The largest advancement of 97.2 ns is observed and exceeds the propagating time of ~50 ns through 10-m SMF. This means that the Stokes light also performs superluminal propagation. Here, the Stokes power in the ring cavity is as 10 times as the output lasing power, which is determined by the optical coupler ratio.

According to Kramers-Kronig relations, the Stokes wave should be slow light, and its delay should increase with the pump power increasing. However, the observed result is opposite to this. To understand the origination of this surprising phenomenon, we change the coupler ratio for 2:8 and choose a shorter fiber of 8-m SMF to observe the slow-light effect of Stokes light. The Stokes power within the cavity and the cavity loss are determined by the coupler ratio. The coupler ratio of 2:8 makes a higher cavity loss and a lower Stokes power which postpones the fast light effect on pump light. If the fast-light effect on Stokes light results from the signal conversion of the superluminal pump signal, we believe that the Stokes wave should perform slow light at the beginning of lasing.

It is reasonable that the pump signal is advanced in time as the pump power is increased. Without fast-light effect, the time delay after passing through an 8-m SMF is ~40 ns. In Fig 3, the advancement of 110.45ns exceeds the normal propagating time and reaches superluminal propagation at negative group velocity which provides a precondition to observe the superluminal Stokes wave.

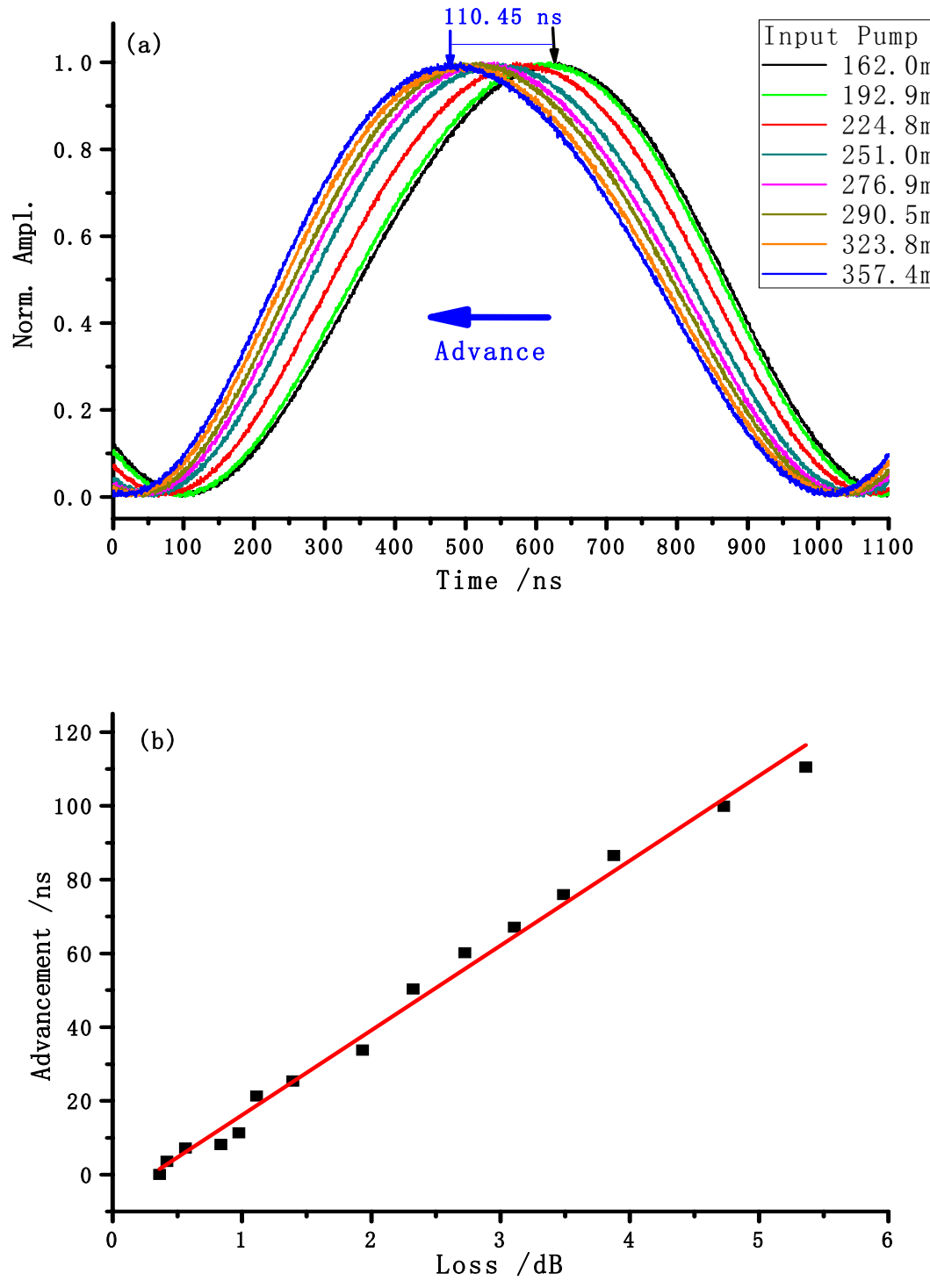

FIG. 3. (a) Waveforms of output pump signal after propagating through an 8-m SMF. The maximum advancement is 110.45ns. (b) The advancement as a function of the loss through the SMF. The advancement varies linearly with the Brillouin loss, and the results show a linear fit with slopes of 23 ns/dB.

In Fig. 4, it is clearly observed that the Stokes wave propagates as slow light when the input pump power range from 186.7 mW to 205.4 mW. In this range, the delay of Stokes light increases with the pump power. However, the Stokes wave advances when the pump power exceeds 205.4 mW. The largest advancement of 93.34 ns is observed and has exceeded the round-trip propagating time (~40 ns) through the 8-m fiber resonator, which means that the Stokes wave performs negative group velocity superluminal propagation as well as the pump light.

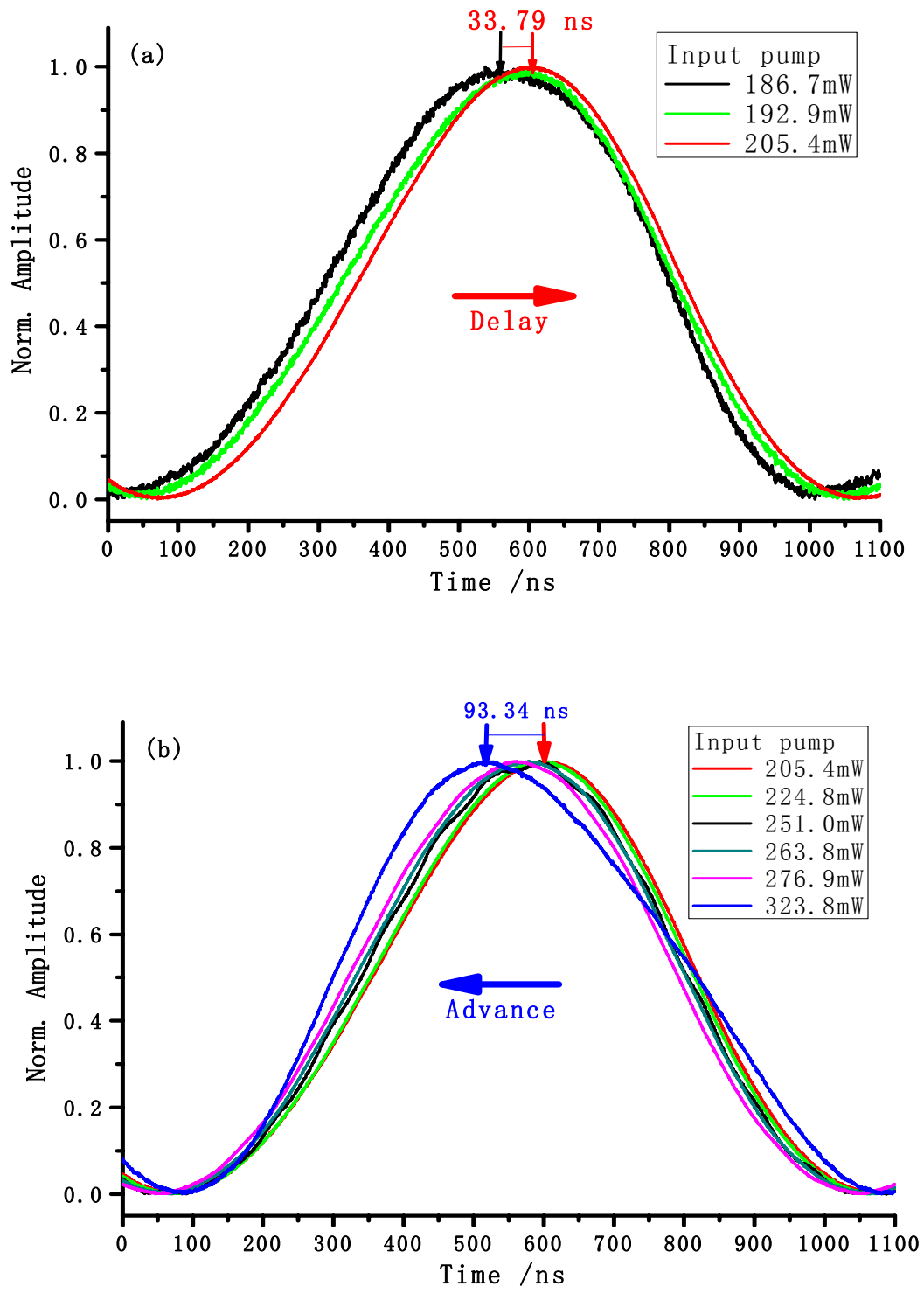

FIG. 4. Waveforms of generated Stokes wave after anticlockwise passing through an 8-m SMF. (a) Slow-light effect of Stokes wave at the beginning of SBS process. (b) Fast-light effect of Stokes wave when the pump power exceeds 205.4 mW.

In typical SBS process [3, 14], the pump and Stokes wave perform fast and slow light respectively, as shown in Fig. 5(a). However, for its low efficiency, the velocity change of both pump and Stokes wave is small compared with the light speed. In addition, the Stokes power is always lower than the pump power.

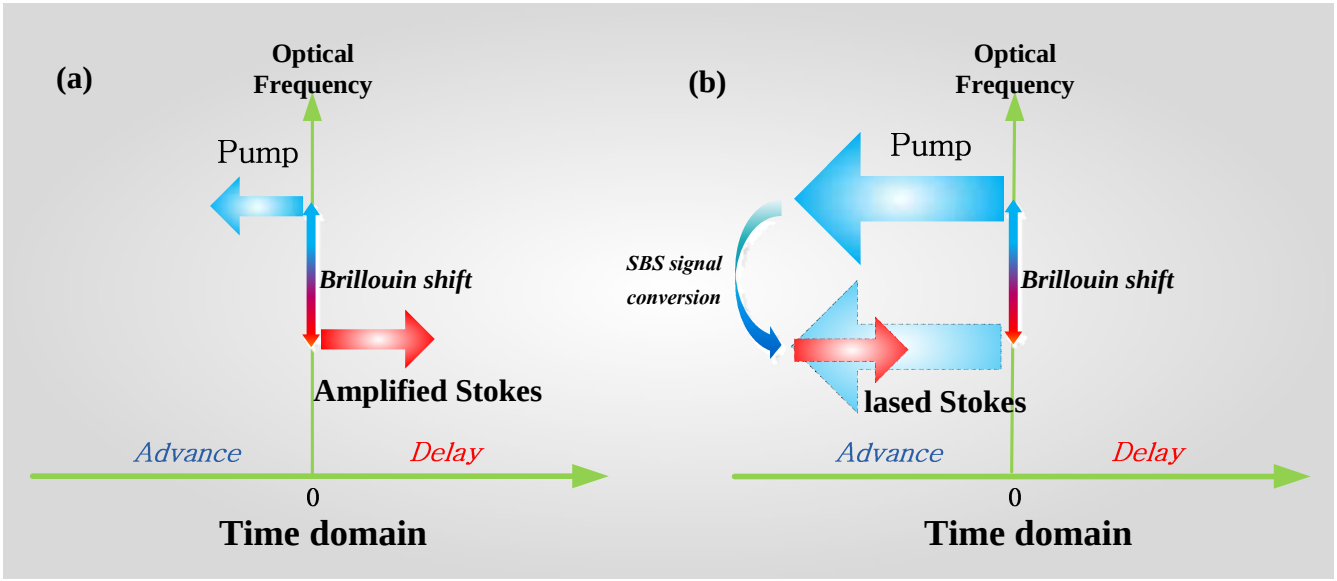

FIG. 5. Schematic illustration. (a) Fast and slow light in typical SBS amplifier system. (b) Superluminal light propagation of pump and generated Stokes wave in the Brillouin lasing resonator.

Nevertheless, the advancement of pump light is greatly enhanced but it is low loss when passing through a Brillouin lasing oscillator [20]. The seeded Stokes pulse always obtains a gain and exhibits slow-light effect. Here, the Stokes signal is converted from the modulated pump signal. When the pump light propagates with a group velocity that is larger than *c*, the generated Stokes pulse peak may exit the medium even earlier than the initial seed pulse peak for the group velocity transition. Superluminal signal conversion from the pump signal results in the fast-light effect on Stokes light. The Stokes light compounds these two effects, and finally may exhibit fast light or even superluminal effect, as shown in Fig 5(b). Importantly, the Stokes power in the lasing cavity can exceed the pump power, which is very different from the conventional SBS process.

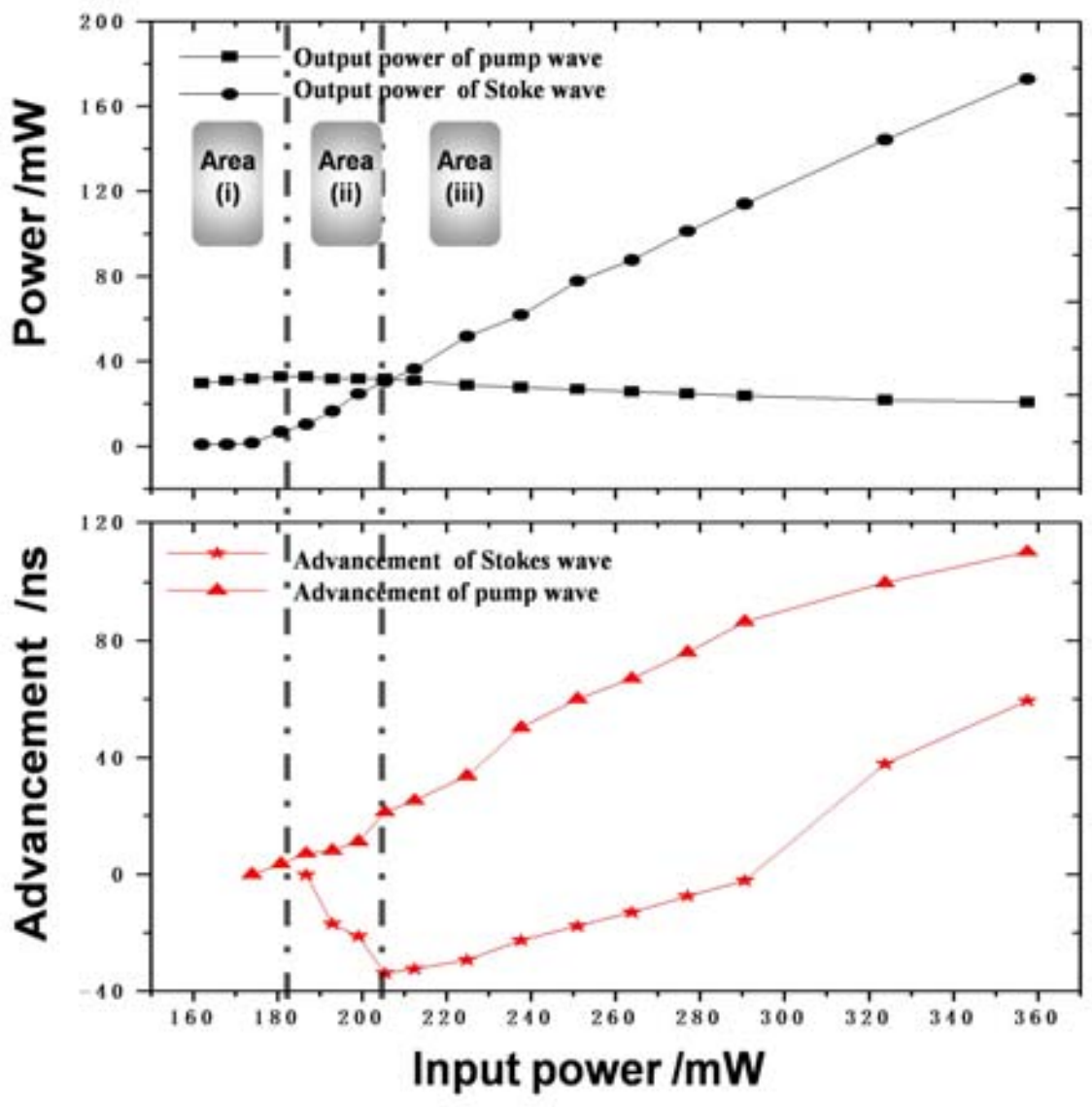

FIG. 6. Output pump/Stokes power versus different input pump powers and the advancement of output pump and Stokes wave as a function of input power. The pump level is divided into three areas: (i) no SBS effect area (below the threshold); (ii) SBS slow light area (the Stokes power in the cavity is lower than the pump); (iii) SBS fast light area (the Stokes power in the cavity is higher than the pump). Correspondingly, the slow/fast light turn point appears at the turn point of power level between pump and Stokes light. The Stokes power in the cavity is as 4 times as the output lasing power. Above the SBS threshold of 186.7 mW, the output Stokes lasing power increases rapidly and then exceeds the output pump power, which means the Stokes power is larger than the pump power within the resonator.

In principle, the SBS process is governed by the coupled mode equations [25]. The Stoke wave exits the fiber with a delay time after passing through a length $L$ of the fiber. When $\delta << 1$, the relative delay time can be written as

$$\Delta T_d = Ln_g / c - Ln_{g0} / c = \frac{Lg_p P_p}{\Gamma_B A_{eff}} . \qquad (2)$$

SBS fast-light for pump is mathematically described as the same as the case for Stokes light, but the only difference is the negative sign of Brillouin gain in the coupling wave equation. Thus, the relative advanced time $\Delta T_a$ for the pump wave after propagating through the fiber is

$$\Delta T_a = Ln_g / c - Ln_{g0} / c = \frac{Lg_p P_s}{\Gamma_B A_{eff}} . \qquad (3)$$

Considering the generation of Stokes wave and the group velocity change of pump light in the SMF, the Stokes wave also obtains a time boost as the pump light. Based on a basic assumption that the optical interaction is determined by the group velocity even at the negative group velocity superluminal propagation, the time delay or advancement $\Delta T_s$ of Stokes light is determined by

$$\Delta T_s = \Delta T_a + (-\Delta T_d) = \frac{Lg_p P_s}{\Gamma_B A_{eff}} - \frac{Lg_p P_p}{\Gamma_B A_{eff}} = \frac{Lg_p}{\Gamma_B A_{eff}}(P_s - P_p) \quad (4)$$

Corresponding to this assumption, the combined slow- and fast-light effects are observed, as shown in Fig. 6. When the pump power is below the SBS threshold, there is no Stokes light. As the input power exceeds the threshold of 186.7 mW, the generated Stokes light obtains a Brillouin gain at first and performs slow light. The slow/fast light turnning point appears at the reversed point of the power level of pump and Stokes light, which confirms with Eq. (4). The Stokes wave performs slow light effect because of the gain, and the maximum delay is 33.79 ns when the input power is 205.4 mW. As the pump power is increasing, the Stokes wave turns to advance in the time domain. With the 323.8-mW pump power, the 93.34-ns advancement is beyond the ~40-ns transmission time of the 8-m SMF, which means the Stokes light propagates with a negative group velocity.

Clearly, the Stokes power results in superluminal propagation of pump light. Then, the pump light in turn converts the superluminal group velocity to the Stokes via SBS process. However, in usual SBS amplifier structure, the Stokes power is always lower than the pump one which means the fast-light advancement of Stokes is smaller than delay time. However, the Stokes power in a lasing resonator can exceed the pump power, which is essential in observing the fast light effect of Stokes wave. This proves that the superluminal Stokes light accompanies the superluminal propagation of pump light when the pump signal converts into the Stokes light.

Besides, we insert two bidirectional 1:99 optical couplers in the Brillouin laser resonator (at point A and B in Fig. 1) in order to measure the group velocity and energy flow in different points within the superluminal propagation. It's observed that the peak of a superluminal pulse propagates in backward at a negative group velocity. Even though energy flow is always forward signal of pump light also converts to the opposite-propagating Stokes light within the optical fibers. Moreover, all these effects are initiated by the far leading edge of the pulse and are consistent with the causality.

Finally, we note that the mechanism of the observed superluminal light via a Brillouin fiber ring resonator differs from the previous slow/fast-light experiments associated with a Brillouin amplifier. In the present experiments, the lased Stokes light within the cavity makes the group velocity of pump wave a tremendous change and superluminal propagation. Combined with the SBS slow-light effect and optical interaction, the Stokes wave performs from slow light to fast and superluminal propagation by increasing the input pump power. The results prove the first evidence that the optical interaction is determined by the group velocity even at the negative group velocity propagation.

The Stokes waves are generated from the pump waves with the wavelength conversion of ~10-GHz frequency shift. It implies that the signal conversion can be realized at the negative group velocity superluminal propagation and the possibility of superluminal information interchange. To date, ultra dense wavelength-division-multiplexed (DWDM) optical transmission around 10-GHz spacing [26-27] has been reported. The development of ultra DWDM makes the possibility of SBS superluminal signal conversion. Also, it allows large superluminal effects that lead to a new examination of the superluminal physics implied by special relativity. We believe that it may play an important role in the implementation of all-optical communication system, interconnection in super computer, optical buffering, high sensitivity sensing, and data synchronization.

This work was supported by the National Natural Science Foundation of China (Grants 61178014/11274231), and the key project of the Ministry of Education of China (Grant 109061).